\documentclass[onecolumn,nofootinbib]{revtex4}

\usepackage{amsfonts,amssymb,amsmath,amsthm,amscd}
\usepackage{graphicx}
\usepackage{dcolumn}
\usepackage{hyperref}
\usepackage{bm}
\usepackage{enumerate}

\def\demi{\frac{1}{2} } 
 \newcommand{\dr}{\rightarrow}   
\newcommand{\R}{\mathbb{R}}   

    \def\lll{{\cal L}}  \def\nn{{\nonumber}}

\def\ka{{\kappa}} 
\newcommand{\be}{\begin{equation}}
\newcommand{\bee}{\begin{equation}}
\newcommand{\eeq}{\end{equation}} 

\def\ov{\overline}

 \newcommand{\bem}{\begin{bmatrix}} \newcommand{\eem}{\end{bmatrix}}
\newcommand{\beq}{\begin{equation}} \newcommand{\ee}{\end{equation}} \newcommand{\eq}{\end{equation}}
\newcommand{\beqa}{\begin{eqnarray}} \newcommand{\eeqa}{\end{eqnarray}}
\def\ka{\kappa}
\def\ie{{i.e. \/}}
\def\eg{{e.g. \/}}

\def\mn{{\mu\nu}}

\begin{document}
\title{Is the notion of time really fundamental?\footnote{{\it Fourth prize at the essay competition of the Foundational Questions Institute }(\texttt{www.fqxi.org}) {\it on The Nature of Time.}}}

\author{F. Girelli} 
\email{girelli@physics.usyd.edu.au}
\affiliation{SISSA, Via Beirut 2-4, 34014 Trieste, Italy and INFN sezione di Trieste \\ and School of Physics, The University of Sydney, Sydney, New South Wales 2006, Australia}
\author{S. Liberati, L. Sindoni}
\email{liberati@sissa.it, sindoni@sissa.it}
\affiliation{SISSA, Via Beirut 2-4, 34014 Trieste, Italy and INFN sezione di Trieste}

\bigskip

\begin{abstract}
From the Physics point of view, time is now best described through General Relativity, as part of space-time which is  a dynamical object encoding gravity. Time possesses also some intrinsic irreversibility due to thermodynamics,  quantum mechanical effects... This irreversibility can look puzzling since time-like loops (and hence time machines) can appear in General Relativity (for example in the G\"odel universe, a solution of Einstein's equations). We take this apparent discrepancy as a warning bell pointing to us that time as we understand it, might not be fundamental and that  whatever theory, lying  beyond General Relativity, may not include time as we know it as a fundamental structure. We propose therefore,  following the philosophy of analog models of gravity, that time and gravity might not be fundamental per se, but only emergent features. We illustrate our proposal using a toy-model where we show how the Lorentzian signature and Nordstr\"om gravity (a diffeomorphisms invariant scalar gravity theory) can emerge from a timeless non-dynamical space.

\end{abstract}

\maketitle

\section{Time and Gravity}\label{metric}

Time has been a mystery to mankind for thousands of years, surely since the very start of natural philosophy. Understanding the nature of Time seems indeed fundamental to understand Reality. Even questioning  about this innate perception of a continuous ``becoming" of things can be addressed from many different approaches:   biology,  psychology and of course physics. In this context, the understanding of time  has evolved a lot since the Greeks\footnote{For example, for Aristotle, time is not fundamental but only a derived concept from the notion of space and motion \cite{barbour}.}, passing from being an absolute (observer independent) characteristic  of space-time as in Galilean relativity to a observer dependent entity in Special Relativity. Finally, the common modern point of view on Time is mostly based on Einstein's theory of General Relativity (GR).

More precisely, in the early XXth century, following the works by Minkowski and Einstein,  Time has been unified with Space into the concept of space-time. The latter is geometrically described by   a (four dimensional) manifold, that is, a topological space equipped with a differential structure, together with a metric $ds^2=g_{\mu\nu}dx^\mu dx^\nu$ \cite{wald}. In all the metric theories of gravity (in particular General Relativity), the metric describes the dynamical degrees of freedom encoding gravity (although in metric affine theories the connection also plays a role in the definition of the gravitational field).

A metric is characterized by a signature, that is the number of its positive and negative eigenvalues\footnote{For non-metric theories, the signature can be tracked back in the inner product in the tangent space. The signature is usually not a dynamical object, i.e. it is a background structure.}. We shall use here the terminology Lorentzian for the case $(-1,+1,+1, +1)$ and Riemannian for the one $(+1,+1,+1, +1)$.  A Lorentzian signature allows one to define space-like ($ds^2>0$), light-like ($ds^2=0$) and time-like distances ($ds^2<0$). A time-like geodesic  is interpreted as the trajectory of a massive test particle moving in space-time. Distances along this geodesic  are usually interpreted as the proper time of the particle, that is, the time measured by a clock associated to the particle.  In the case of a Riemannian signature $(+1,+1,+1, +1)$, one can not then talk about time at all since we have only space-like distances ($ds^2>0$).
The notion of signature is therefore an essential part of the notion of ``time".

So far we have considered the textbook definition of time in metric theories of gravity.  However the situation is clearly more subtle than that. Indeed, physical phenomena do possess an intrinsic irreversibility. This irreversibility can have a thermodynamical origin, due to the natural coarse-grained description of Nature.  It  can be also tracked back from the quantum mechanical nature of matter fields (putting aside the issue of quantum gravity for a moment). Indeed the basic axioms of quantum mechanics impose a one-way direction in time evolution: measurement process induces a collapse of the wave-function which cannot be undone going backward in time. Using the words of Ellis \cite{Ellis:2006sq}, there is an ``eternal becoming", which pinpoints not only the time direction, but also its arrow. However, it is also true that when confronted with the most daunting solutions of the Einstein equations this more elaborate intuition about the nature of time seems at least severely challenged: GR does allow for solutions which contains closed time-like loops (e.g. G\"odel universe) and hence time machines (see e.g. \cite{VisserBook}). More generally, time-orientability of space-time does not seem to be a built-in property of the theory.

With respect to these puzzling aspects of General Relativity one can have at least two antithetic points of views. The first, the one most pursued so far, is that time-machines solutions (and other ``time-menacing" features) of GR are just a sign of the limits of the theory, and  that a quantum gravity theory will indeed resolve these issues in pretty much the same way one expects it to remove singularities. On the other hand, one may conjecture that these riddling solutions are just warning bells that our notion of time may not be fundamental; possibly whatever lies beyond General Relativity (and its related notion of space-time manifold) may not include Time as we know it as a fundamental structure\footnote{All in all, it is well known that whenever chronological horizons arise,  quantum field theory in curved space-time also breaks down (Kay-Radzikowsky-Wald theorem~\cite{KRW}), and quantum gravity has to be called in, in order to make any prediction.}.

However, given that even in GR the notion of time as ``time-arrow" is problematic, we shall focuss here on the emergence of time in the sense of Lorentzian signature from a Riemannian one. More specifically, in this short essay we want to discuss the possibility that space-time and its dynamics, e.g. GR, are all emergent at the same level from some timeless fundamental, non-gravitational, theory. In order to do so, however, we shall have to pull back to some simpler gravitational theory than GR, and in particular we shall work with scalar fields and a scalar theory of gravitation, \ie Nordstr\"om gravity  \cite{Nordstrom}.

\medskip

Historically, the  first relativistic version of a gravitational dynamics was proposed by Nordstr\"om, who in 1913 tried to generalize the Poisson equation, while preserving, following Einstein's advice, both what would later be called the strong equivalence principle \cite{Will}, and the intrinsic non-linearity of the gravitational interaction. After some years, Nordstr\"om finally succeeded, and his gravity theory (minimally coupled to a scalar field $\phi$) was later described in a geometric way by Einstein and Fokker  as follows
\begin{eqnarray}
C_{\alpha\beta\gamma\delta}=0,\label{EOM2} \\
\textbf{R}= \ka \textbf{T}, \label{EOM1}\\
 (\square_g + m^2)\phi =0,\label{EOM3}
\end{eqnarray}
where $\textbf{T}$ is the trace of the stress energy tensor $T_\mn$, defined with respect to $g_\mn$, for the matter field $\phi$, $\textbf{R}$ is the Ricci scalar for $g_\mn$ and $\ka$ is proportional to the Newton constant $G_{\rm N}$. Equation \eqref{EOM2} encodes the fact that the Weyl tensor $C_{\alpha\beta\gamma\delta}$ is zero, \ie that the metric $g_\mn$ is conformally flat. Equations \eqref{EOM2} and \eqref{EOM1} are  called the Einstein--Fokker equations\footnote{Note that this almost the same, modulo a minus sign, as the trace of the Einstein equation restricted to the conformal metric.}.  Albeit much simpler than GR, it is important to stress that  Nordstr\"om gravity  is  a consistent theory of gravitation, and that it shares many features with GR. Most importantly it does satisfy the strong equivalence principle and is diffeomorphisms invariant. It is however clearly not physical since fields which are described by conformally invariant equations of motion will always move as in flat space-time. As such this theory is hence unable to deflect observed phenomena as the bending of light by gravitational fields. Furthermore its diffeomorphisms invariance is obtained at the cost of background independence, given that in this case, in addition to topology and signature, another background structure is introduced, namely the Minkowski metric. Let us nonetheless take this simpler theory of gravity as our target, and show how time and gravity may emerge at the same level.

\section{Is the notion of time really fundamental?}

Before embarking into the specific discussion about the emergence of time, let us make a small detour explaining what ``emergence'' means, starting from the very well understood case of condensed matter.

\medskip

In condensed matter systems, the  macroscopic properties of a system are  determined by the way  in which the microscopical degrees of freedom (\eg the atoms) are organized. However, the micro- and macro- worlds cannot be connected in an easy way: it can be a daunting  task to derive macro equations from micro equations. Conversely, in general it is difficult to derive  the micro-physics from macro-physics.

For example, water is described by hydrodynamic equations. Quantizing these equations will not provide the correct fundamental micro-physics. One needs to be aware of the existence of atoms to be able to propose a satisfying fundamental theory for water. On the other hand it is a difficult task to obtain the hydrodynamic equations from the Schr\"odinger equation encoding the dynamic of the hydrogen and oxygen atoms. This example illustrates  how the fundamental dynamics (\ie here the Schr\"odinger equation) can be qualitatively different than the ``emerging" dynamics (\ie here the hydrodynamic equations). It emphasizes also how the macroscopic or ``emergent" degrees of freedom (fluid) have different properties than the microscopic degrees of freedom (atoms).

\medskip

Roughly speaking, an ``emergent theory"  is obtained by proceeding to a large large $N$ limit for some class of degrees of freedom in the fundamental theory. It is often the case that the resulting theory shows properties qualitatively different than the ones of the fundamental theory. A typical way to construct such emergent theory is to condense all the relevant properties in some macroscopical/coarse-grained  variables which somehow encode the universal properties of the large number system without entering into the details of the microscopic dynamics. While the detailed description of macroscopic dynamics would be only approximate, one can still discuss some of its essential features.
For instance, a Bose-Einstein condensate is described by a nonlinear Schr\"odinger equation for a  classical complex scalar field, rather than from the n-body quantum state describing all the atoms in the condensate. Despite this  very simple description, the phenomenon of condensation and the properties of the quasi-particles are correctly predicted. Note that this description is only an approximation, valid  only in a given regime. When getting out of the regime, one has to take into account the fundamental theory instead of the emerging theory.

\medskip

One could expect a similar situation to happen in the context of Quantum Gravity\footnote{We assume here that  the fundamental theory behind gravity is obtained by quantization of the relevant degrees of freedom. This assumption could be wrong if the quantum theory is {\emph{per se}} an emergent notion too, see for instance \cite{smolinemqm} and references therein.}. Einstein equations could be the analogue of some type of hydrodynamic equations. Even if a Quantum Gravity theory was fully available, the derivation of Einstein equations from the fundamental equations of Quantum Gravity would then be something extremely nontrivial, such as in the case of water for example. Moreover, this would indicate also  that the fundamental Quantum Gravity theory has probably nothing to do with (quantum) geometry. The latter -- and its dynamics -- would emerge only in some type of macroscopic limit.

In fact,  there exist various models, called ``analogue models for gravity", where a Lorentzian metric appear in a suitable regime, even though at the fundamental level this geometrical structure is absent \cite{stefano}. A well studied example is again the Bose-Einstein condensate,  where it was shown that phonons, \emph{in some regime}, propagate in some (curved) Lorentzian geometry. In this sense, the Lorentzian geometry is emerging: it is not a description of the fundamental physics, it is only a valid description of the physics in a given regime, for some degrees of freedom.

Phonons are interpreted as (massless) matter in this model. In some sense they can be also interpreted as emerging degrees of freedom. Indeed, a phonon is a collective degree of freedom constructed from the atoms outside the condensate. Typically the fraction of atoms which are not condensed is of order $1/N$, where $N$ is the number of atoms in the condensate. When $N$ is large that is the condensation is happening, the phonon can  be considered as a perturbation around the condensate. In this sense, perturbations can be seen as coming from some large $N$ limit and can encode ``emerging degrees of freedom".

These analog models provide therefore some interesting examples where both matter and the metric do emerge. Unfortunately, there is a priori no known construction to obtain the Einstein equations, encoding the dynamics for the metric (albeit  in Bose-Einstein condensate it is possible to recover in some limit a Newtonian-like dynamics for the background \cite{lorenzo}).

Although, still yet uncomplete, these models provide therefore some interesting relationships between geometrical structures and condensed matter systems illustrating the concept of emergence. There are also some striking relationships of Gravity with thermodynamics (\ie again some large $N$ limit). We can cite for example the well-known  black hole thermodynamics or the derivation of the Einstein equations from a thermodynamical state equation by Jacobson \cite{jacobson}. These results indicate that General Relativity could be an emergent theory, arising from some nontrivial large $N$ limit from a Quantum Gravity theory that does  not have to be necessarily related to the notion of geometry.

\medskip

The most developed Quantum Gravity theories at this stage are String theory and Loop/Spinfoam quantum gravity. The first considers roughly that the fundamental degrees of freedom describing gravity are encoded in a string. In this sense, this could be considered as  an example of emergent gravity since the macrophysics (General Relativity) is very different than the microphysics (string physics).  The Loop/Spinfoam approach consists in a direct (canonical or by path integral) quantization  of  General Relativity using a non-metric formulation. In this context, it is clearly assumed that the fundamental Quantum Gravity theory is about quantum geometry\footnote{ There are recent results corroborating this approach where discretized gravity is obtained after a semi-classical limit from some spinfoam amplitude \cite{conrady}. Note however that there are also some proposals to obtain General Relativity as some large (thermodynamical/statistical) $N$ limit from spinfoams encoded in some type of (Group) Field theory \cite{oriti}.}. We are however interested to know if the features describing time that we listed in the previous section, are really fundamental, that is associated to the fundamental Quantum Gravity theory. In both Quantum Gravity theories mentioned above, the signature is, in general, a background structure, that is \emph{not} dynamical and hence, following the above discussion, these are theories with a built-in time notion which at most can resolve the above mentioned paradoxes in GR by forbidding them rather that replacing or eliminating the notion of Time. As previously said, our aim here is different in the sense that we shall show instead that signature and a diffeomorphisms invariant theory of gravitation can both be emergent concepts, not built in the fundamental theory.

\section{A toy-model}
In the following toy-model \cite{lorenzo1}, we shall consider, largely following the intuition gained from condensed matter analogues of gravity \cite{stefano}, the possibility that Lorentzian dynamics can emerge as a property of the equations associated to perturbations around some solutions of the equations of motion. These perturbations will be associated to the emergent gravity and matter degrees of freedom which will characterize our gravitational dynamics.

\subsection{Description of the toy-model}
Consider the (quantum) fundamental theory underlying space-time which could be some type of graph theory, string theory... The exact details of this theory is not of interest for our purpose. Following the inspiration of the analog models for gravity, we assume that from this theory, there is some type of ``condensation", so that the condensate is described by a manifold\footnote{We are not interested here on how a manifold can emerge, though it is of course an interesting question to explore.} $\R^4$  equipped with the Euclidean metric $\delta^\mn$. There is therefore no Lorentzian structure, both the condensate and the fundamental theory are \emph{timeless}. We assume also that out of this condensation process, a set of scalar fields $\Psi_i(x_\mu)$, $i=1,2, 3$ together with their Lagrangian $\lll$  are emerging. The number of scalar fields is not really relevant here, we only need to have at least two fields for our construction. $\lll$  is invariant under the Euclidean Poincar\'e group $ISO(4)$ and of the general shape
\begin{equation}\label{initial lagrangian} {\lll=F(X_1, X_2, X_3)}= f(X_1)+f(X_2)+f(X_3), \qquad  X_i=\delta^{\mn}\partial_\mu\Psi_i\partial_\nu\Psi_i.\end{equation}
As an example we can take
\be\label{example of lagrangian}f(X)= -X^2+X= -(\delta^{\mn}\partial_\mu\Psi\partial_\nu\Psi)^2 +\delta^{\mn}\partial_\mu\Psi\partial_\nu\Psi. \ee
The equation of motion for the field $\Psi_i(x_\mu)$ are simply given as
\begin{equation}\label{EOMf}
 \partial_\mu \left(\frac{\partial F}{\partial X_i}\partial^\mu\Psi_i\right)=0=\Sigma_j\left(\frac{\partial^2F}{\partial X_i\partial X_j}(\partial^\mu X_j)\partial^\mu\Psi_i + \frac{\partial F}{\partial X_i}\partial_\mu\partial^\mu\Psi_i\right).
 \end{equation}

As said previously, we want to focus on the perturbations $\varphi_i$ around solutions $\psi_i$ of the above equation, that is such $$\Psi_i=\psi_i+\varphi_i.$$
As we shall see below, the perturbations will encode both the gravitational and matter degrees of freedom.  $\varphi_i$ as a perturbation is the analog of the phonon in the Bose-Einstein condensate. In this sense they can be considered as emerging degrees of freedom.

The Lagrangian for the $\varphi_i$ is constructed by considering the expansion of $F(\ov{X}_i+\delta X_i)$, with
$$ \ov{X}_i= \delta^{\mn}\partial_\mu\psi_i\partial_\nu\psi_i \quad \textrm{ and } \quad \delta X_i= 2\partial_\mu\psi_i\partial^\mu\varphi_i +\partial_\mu\varphi_i\partial^\mu\varphi_i.$$
It is explicitly given by
\be\label{lagrangian for perturbation}\begin{array}{c}
F(X_1, X_2, {X}_3)\dr F(\ov{X}_1, \ov{X}_2,\ov{X}_3 )+ \sum_j\frac{\partial F}{\partial X_j }(\ov{X})\delta X_j + \demi \sum_{jk}\frac{\partial^2 F}{\partial X_j\partial X_k }(\ov{X})\delta X_j\delta X_k \\ + \frac{1}{6}\sum_{jkl}\frac{\partial^3 F}{\partial X_j\partial X_k \partial X_l}(\ov{X})\delta X_j\delta X_k\delta X_l+ ...\end{array}\ee
We are ready now to see what are the necessary conditions to put on the classical solutions $\psi_i$ such that the perturbations $\varphi_i$ do effectively see a Lorentzian metric.

\subsection{Emerging the Lorentzian signature}
To determine the effective metric in which the perturbations $\varphi_i$ propagate, we need to look at the kinematic term, which will depend on the solution $\psi_i$. In fact different choices of solution $\psi_i$ will lead to different metrics. From \eqref{lagrangian for perturbation}, we determine explicitly the tensor $g^\mn$ standing in in front of $\partial_\mu \varphi_k\partial_\nu \varphi_k$, $\forall k=1,2, 3$:
$${g^\mn_k \equiv \frac{d f}{d X_k }(\ov{X}_k)\delta^{\mn} + \demi \frac{d^2 f}{(d X_k)^2 }(\ov{X}_k)\partial^\mu \psi_k\partial^\nu \psi_k.}
$$
We note that the metric can be different for different fields: according to the choice of $\psi_k$, the metric seen by the field $\varphi_k$ can be different than the metric seen by the field $\varphi_{k'}$ for some choices of different solutions $\psi_k$ and $\psi_{k'}$. This approach can lead naturally to a multi-metric structure. In order to have mono-metricity, we need to choose similar solution for each of the field, so that $\psi_k=\psi,$ $\forall k$.

Let us now consider a specific class of solutions of our equations of motions (\ref{EOMf})
\be\label{solution}\psi_i=\alpha_i ^\mu x_\mu +\beta_i.\ee
The constants $\alpha_i$ and $\beta_i$ are encoding the boundary conditions and following the above discussion we shall take the case in which they are field independent:
${\psi}= \alpha^\mu x_\mu + \beta$.

Thanks to the $SO(4)$ symmetry, we can always make a rotation such that
\be\label{choice of psi} \bar{\psi}= \alpha x_0 + \beta.\ee
The choice of the coordinate $x_0$ is completely arbitrary, what only matters is that there is one coordinate which is pinpointed. For such choice the effective metric takes the shape
\be
g_{00}=  \frac{d f}{d X }(\ov{X})+\frac{\alpha^2}{2}\frac{d^2 f}{(d X)^2 }(\ov{X}), \qquad g_{ij}= \frac{d f}{dX}(\ov{X})\delta_{ij}, \qquad g_{\mu\nu}=0 \quad \forall \mu\neq\nu.\nn
\ee
In order to get the Lorentzian metric we see that we need to have the following conditions between the derivatives of $f$ and the initial condition $\alpha$:
\be{\frac{d f}{d X }(\ov{X})+\frac{\alpha^2}{2}\frac{d^2 f}{(d X)^2 }(\ov{X})<0, \qquad   \frac{d f}{d X }(\ov{X})>0}.\nn
\ee
When $f$ is for example ${f(X)= -X^2+ X}$, we just need to have $\alpha$ such that
\be\label{constraint for alpha}\frac{1}{2}<\alpha^2<\frac{1}{3}.\ee
Modulo a (constant) rescaling of the coordinates $x_\mu$, we can then obtain the Minkowski metric, and the perturbations $\varphi_i$ are propagating effectively on the Minkowski metric $ \eta_{\mn}=\mbox{diag}(-1,+1,+1,+1,+1)$.
\be{\lll_{\rm eff}(\varphi_1,\varphi_2, \varphi_3 )= \sum_i  \eta^{\mn} \partial_\mu\varphi_i\partial_\nu\varphi_i.}\ee
The Minkowski metric and the Lorentzian signature  are therefore naturally emerging.

Before discussing the gravity case, let us pause a moment and discuss  various aspects of this result.
So far, our theory does not posses any fundamental speed scale. This is natural since the fundamental theory is Euclidean. At this level, there is no coordinate with time dimension and therefore one cannot define a constant with speed dimension. The invariant speed $c$, which will relate the length $x_0$ to an actual time parameter $t$, could be determined experimentally by first introducing a coordinate with time dimension (as it would be natural to do given the hyperbolic form of the equations of motion for the perturbations)  and then by defining $c$ as the signal speed associated to light cones in the effective spacetime\footnote{
Noticeably, a similar situation is encountered in the von Ignatowsky derivation of Special Relativity \cite{ignatowsky} where, given a list of simple axioms, one derives the existence of a universal speed, observer independent, which is not fixed a priori to be the speed of light but has to be identified via actual experiments.}.

It is clear by construction that there is a lack of criterion for  selecting the right solutions to obtain the Lorentzian signature. For example, one could obtain as well a Euclidean metric if $f$ is as chosen above and $\alpha$ does not satisfy the constraint \eqref{constraint for alpha}. Some coordinate dependent metric  could also be obtained if some non-linear solution $\psi$ could be found.
Normally we select states in physical systems labeling them by means of energy.
Indeed the notion of energy, besides the Lagrangian, plays a role in defining the properties of the physical systems.  Since we are dealing with a system which is fundamentally Euclidean, the notion of energy looses its meaning. In order to have the right solution  naturally selected,   we could add another thermodynamical potential to be minimized. Clearly, this is not unique, and we are trading the problem of time with the problem of finding the right potential.

In fact, while it is conceivable that in a more complicate model we could have some mechanism for selecting the specific background
solution that leads to an emergent time, it is not obvious at all that such a feature should be built in the emergent
theory. Indeed, it is conceivable that the actual background solution in which the initial system of fields \eqref{initial lagrangian} emerges
from the fundamental (pre-manifold) theory, can be determined from the conditions within which the "condensation"
of the fundamental objects takes place. To use an analogy, the same fundamental constituents, e.g. carbon atoms,
can form very different materials, diamond or graphite, depending on the external conditions during the process of
formation. Alternatively, in a Bose-Einstein condensation the characteristics of the background solution (the classical
wave function of the condensate), such as density and phase, are determined by physical elements (like the shape of
the electromagnetic trap or the number and kind of atoms involved) which pre-exist the formation of the condensate.

\begin{figure}
\includegraphics[scale=.7]{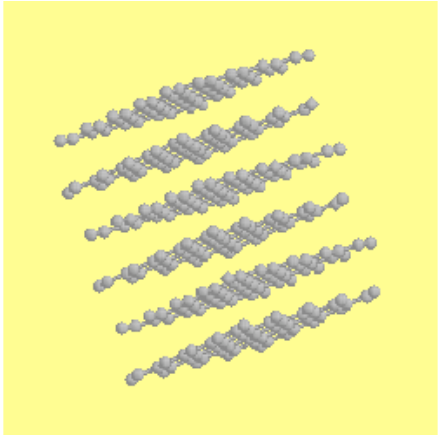} $\qquad \qquad \qquad \qquad \qquad \qquad$
\includegraphics[scale=.7]{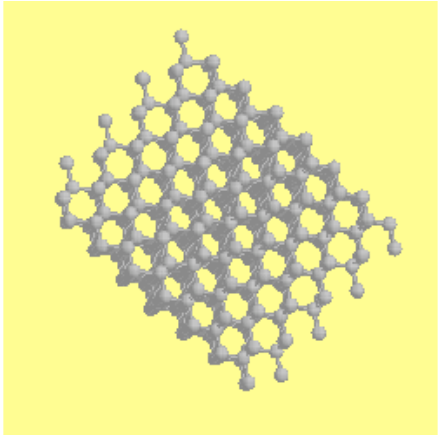}
\caption{On the left, the graphite lattice has a preferred slicing, contrary to the diamond lattice on the right \cite{brisbane}.}
\end{figure}

\subsection{Emerging a diffeomorphisms invariant space-time dynamics}
In the previous subsection, we showed how some degrees of freedom could propagate in a Lorentzian manifold in some regime, even though the fundamental theory is Euclidean. To really have the full notion of (space-)time emerging we need now to see how one can obtain some kind of dynamical space-time coupled to the matter degrees of freedom. In particular the dynamics of space-time needs to be diffeomorphisms invariant.

The first step to do so is to  identify the gravitational and matter degrees of freedom within the perturbations encoded in the multiplet  $\varphi= (\varphi_1,\varphi_2, \varphi_3)$ associated to the simple Lagrangian\footnote{It is invariant also under the global group $O(3)$.} $\lll_{\rm eff}(\varphi)$ that we just obtained above.
\be\label{free lagrangian 1}
\lll_{\rm eff}(\varphi)= \eta_\mn (\partial^\mu  \varphi)^T (\partial^\nu  \varphi).\ee
In order to make apparent the different nature of the degrees of freedom, we  proceed first  to a   change of variables and then analyze the new shape of the equations of motion, expressed in terms of these new variables.
\be{\left(\begin{array}{c}\varphi_1\\\varphi_2\\\varphi_3\end{array}\right)= \Phi\left(\begin{array}{c}\phi_1\\ \phi_2\\ \phi_3\end{array}\right)},  \qquad \textrm{with } {|\phi|^2\equiv\sum_i \phi_i^2=\ell^2}.\ee
Here, for the moment, $\ell$ is just an arbitrary length scale to keep the dimension right.

Under this change of variable, the Lagrangian \eqref{free lagrangian 1}  becomes then
\be\label{new free lagrangian}{\int dx^4 \lll_{\rm eff}(\varphi_i )\dr \int dx^4 \left(\ell^2 \eta^{\mn} \partial_\mu\Phi\partial_\nu\Phi + \sum_i\Phi^2 \eta^{\mn}\partial_\mu\phi_i\partial_\nu\phi_i + \lambda(|\phi|^2-\ell^2)\right)}\ee
$\lambda$ is a Lagrange multiplier encoding the constraint $|\phi|^2=\ell^2$. The equations of motion are easy to derive. We perform the variations over $\phi_i$, $\Phi$ and $\lambda$ respectively:
\begin{eqnarray}
&& \eta^{\mn} (2\partial_\mu\Phi\partial_\nu\phi_i + \Phi^2 \partial_\mu\partial_\nu\phi_i + \frac{1}{\ell^{2}}\partial_{\mu}\phi_{j} \partial_{\nu}\phi_{k}\delta^{jk}\phi_{i}) =0, \label{matter EOM 1}\\
&& \eta^{\mn} (\ell^2\partial_\mu\partial_\nu\Phi - \Phi \sum_i\partial_\mu\phi_i\partial_\nu\phi_i) =0,\label{gravity EOM}\\
&& |\phi|^2-\ell^2=0. \label{constraint}
\end{eqnarray}
By introducing the  metric
\be\label{conformal-metric}{g_{\mn}(x)= \Phi^2(x)\eta_{\mn}},\ee
 we can rewrite \eqref{matter EOM 1} simply as \cite{lorenzo1}
$$\begin{array}{c}
{\Box_g \phi_i + \frac{1}{\ell^{2}} g^{\mu\nu}\partial_{\mu}\phi_{j}\partial_{\nu}\phi_{k}\delta^{jk}\phi_{i}=0} .
\end{array}$$
We recognize the equation of motion for a nonlinear sigma model, propagating on the (curved) space-time given by the metric $g_{\mn}(x)$. This suggests  that {\emph{the gravitational degree of freedom should be encoded in the scalar field} $\Phi$}, whereas {\emph{matter should be encoded in the} $\phi_i$}.

One should be careful in the counting of degrees of freedom. Indeed the metric  \eqref{conformal-metric} is conformally flat, that is the Weyl tensor is zero $
{C_{\alpha\beta\gamma\delta}(g)=0}$. In order to keep the same number of degrees of freedom, \eqref{matter EOM 1} should be really rewritten as
$${ \eta^{\mn} (2\partial_\mu\Phi\partial_\nu\phi_i + \Phi^2 \partial_\mu\partial_\nu\phi_i+\frac{1}{\ell^{2}}\partial_{\mu}\phi_{j} \partial_{\nu}\phi_{k}\delta^{jk}\phi_{i})  =0} \longrightarrow  \left(\begin{array}{l} \Box_g \phi_i+\frac{1}{\ell^{2}}g^{\mu\nu}\partial_{\mu}\phi_{j} \partial_{\nu}\phi_{k}\delta^{jk}\phi_{i}=0 \\ C_{\alpha\beta\gamma\delta}(g)=0. \end{array}\right.$$
Thanks to these insights, we come back to the matter contribution in the action \eqref{new free lagrangian}, and we rewrite in terms of the metric variable $g_\mn$.  Recalling that $ \sqrt{-g}=\Phi^4$ and $g^{\mn}= \Phi^{-2}\eta^{\mn}$
$$ \begin{array}{c}\int dx^4 \left( \sum_i \Phi^2 \eta^{\mn} \partial_\mu\phi_i\partial_\nu\phi_i + \lambda(|\phi|^2-\ell^2)\right)\longrightarrow \int \left(\sum_ig^{\mn}\partial_\mu\phi_i\partial_\nu\phi_i + \lambda'(|\phi|^2-\ell^2)\right)\sqrt{-g}dx^4.\end{array}$$
Note that we have rescaled the Lagrangian multiplier that became $\lambda'$. We can therefore construct the matter stress-energy tensor $ T_\mn(\phi_i)$   associated to the metric $g_\mn$, as well as its trace $\textbf{T}(\phi_i)$ \cite{lorenzo1}.
\begin{eqnarray}\label{ST}
&& \label{trace ST} {\textbf{T}(\phi_i)= g^\mn T_\mn(\phi_i) = - \Phi^{-2}\sum_{i}\eta^{\mn}\partial_\mu\phi_i\partial_\nu\phi_i}.\nn\end{eqnarray}
 We recognize here that the trace  $\textbf{T}(\phi_i)$  is proportional to a term present in \eqref{gravity EOM}. In fact, recalling that  for the  conformally flat metric \eqref{conformal-metric}, the Ricci scalar $\textbf{R}$ is given as
$${\textbf{R}=  - 6 \frac{\Box_\eta \Phi}{\Phi^{3}}},
$$
the equation of motion \eqref{gravity EOM} becomes after some elementary algebra
\be\label{einstein fokker 0}
{\textbf{R } = \frac{6}{\ell^2}  \textbf{T}}.\nn\ee
If we introduce the gravitational constant $G_N$ and recall that the Weyl tensor is zero, we recognize the  \emph{Einstein--Fokker equations}  \cite{fokker}:
\begin{equation}\label{Nord}
\textbf{R}= 24  \pi G_{\rm N}\, \textbf{T}, \quad C_{\alpha\beta\gamma\delta}=0,
\end{equation}
where $G_{\rm N}$ in our model has to be proportional to $\ell^{-2}$ \cite{lorenzo1}. In this sense we now see that the scale $\ell$ is related to the effective Planck scale of our model. Noticeably, the fact that the underlying fundamental signature is Riemannian allowed us to introduce it without affecting the relativity principle at high energy (a fundamental, i.e. observer independent, length scale is problematic if boost invariance is required at all scales but it is perfectly compatible with the Euclidean Poincar\'e group that we are imposing on our starting emergent manifold).

To summarize, we showed that under some change of variable, the free Lagrangian for a family of scalar fields can be precisely rewritten as a non-linear sigma model coupled to Nordstr\"om gravity.
$$\left.{\begin{array}{c}
 \eta^{\mn} (\partial_\mu\partial_\nu\Phi - \Phi \sum_i\partial_\mu\phi_i\partial_\nu\phi_i) =0\\
 \eta^{\mn} (2\partial_\mu\Phi\partial_\nu\phi_i + \Phi^2 \partial_\mu\partial_\nu\phi_i+\frac{1}{\ell^{2}}\partial_{\mu}\phi_{j} \partial_{\nu}\phi_{k}\delta^{jk}\phi_{i}) =0 \\
|\phi|^2-\ell^2=0 \end{array}}\right)
 \dr
 \left({\begin{array}{c}
\textbf{R} = \frac{6}{\ell^2}  \textbf{T}   \quad \\
 \Box_g \phi_i+ \frac{1}{\ell^{2}}g^{\mu\nu}\partial_{\mu}\phi_{j} \partial_{\nu}\phi_{k}\delta^{jk}\phi_{i})=0 \\
 |\phi|^2-\ell^2=0, \qquad \quad C_{\alpha\beta\gamma\delta}(g)=0 \end{array}}\right.
$$
We have shown how the equations \eqref{matter EOM 1}-\eqref{gravity EOM} can be rewritten in an evidently diffeomorphisms invariant form, from the point of view of ``matter fields observers" (the $\phi_i$). Following the standard hole argument (see \eg \cite{sonego}), this also implies that the coordinates $x_\mu$, used to parameterized our theory, do not have any physical meaning from the point of view of the $\phi_i$ ``matter observers". They are merely parameters. The diffeomorphisms symmetry is therefore ``emerging". {Of course this theory has more background structure than GR: there is evidently a preferred metric, the Minkowski one, however this is invisible to the matter fields as none of them has conformally invariant equations of motion.}

\section{Discussion: Dynamical space-Time from a timeless non-dynamical space}
We have showed, in a toy model, how the Lorentzian signature and a dynamical space-time can emerge from a flat non-dynamical Euclidean space, with no diffeomorphisms invariance built in. In this sense the toy-model provides an example where time (from the geometric perspective) is not fundamental, but simply an emerging feature. Accidentally this non-fundamental feature of time should automatically remove any issue about causality, as described in the first section of the essay, given that the very notion of causality would cease to be valid at high energies (which are always involved in the presence of chronological horizons \cite{VisserBook}).

The Lorentz  and the diffeomorphisms symmetries are \emph{emerging symmetries} that is only approximate symmetries. Indeed when taking into account the third order terms in \eqref{lagrangian for perturbation}, these symmetries are simply broken. This is quite natural since the model \eqref{initial lagrangian} has nothing to do with either of these symmetries.

\medskip

Since we have a concrete model, one can ask if the observers living in such emergent space-time could foresee that  time is actually not fundamental. There are in fact various possible ways, even though this could be quite difficult experimentally. For example, the first obvious way is to be able to measure a Lorentz symmetry breaking contribution in high-energy (in our case in the form of a non-dynamical ether field).
There could be also a possible window in the strong gravitating regime. Indeed  our derivation obviously holds for small perturbations $\varphi_i$, and hence small $\Phi$, implying that in our framework one would predict strong deviations from the weak field limit of the theory whenever the gravitational field becomes very large. It would be interesting to see how these deviations actually do appear.

\medskip

We want to emphasize again that the toy-model describes Nordstr\"om gravity, which is clearly non-physical since  there is no bending of light. It would be extremely interesting to be able to derive in a similar way General Relativity, even though this seems to be a very difficult task {(for example, it would probably require to have a mechanism that selects only a background signature and not the whole Minkowski metric as in our case, given that only the former is allowed to be non-dynamical in GR)}. One would in this case aim to obtain the emergence of a theory characterized by spin-2 gravitons (while in Nordstr\"om theory the graviton is just a scalar). This could open a door to a possible conflict with the so called Weinberg-Witten theorem \cite{WW}. However, there are many ways in which such a theorem can be evaded (see \eg \cite{Jenkins}) and in particular one may guess that analogue models inspired mechanisms like the one discussed here will generically lead to Lagrangian which show Lorentz and diffeomorphisms invariance only as approximate symmetries for the lowest order in the perturbative expansion (while the Weinberg-Witten theorem assumes exact Lorentz invariance).

The emergence of time in our toy-model is also very much depending on the choice of the solution $\psi$. In fact the specific choice of $\psi$ might even seem a bit contrived.
The fact that we are unable to choose  $\psi$ might be due that other ingredients are needed to fully understand how a time-like direction can emerge in a natural way out of an Euclidean system. This might be due to the fundamental theory, as it has been pictorially described by the diamond/graphite example, or due to some other criterion, like a minimization of another state functional which we presently do not know.
The toy-model would clearly gain in strength if one is able to show  that the solution $\psi$ is preferred for some reason. In this sense one could predict the apparition of time. 

\medskip

We are clearly aware of these two drawbacks (\ie Nordstr\"om is not General Relativity, the choice of $\psi$), however we think that they should not hide our main point: it is possible to construct a toy-model and to identify at least one solution $\psi$ such that a dynamical space-time together with  matter do emerge from a timeless non-dynamical space. We have therefore provided an example where \emph{time and gravity are not fundamental but only emergent}.



\begin{thebibliography}{99}

\bibitem{barbour}
J. Barbour, \emph{The discovery of dynamics}, Oxford University Press (2001).





\bibitem{wald} R. Wald, \emph{General Relativity}, University Of Chicago Press (1984).

\bibitem{Ellis:2006sq}
 G.~F.~R.~Ellis,
 \emph{Physics in the real universe: Time and spacetime},
 Gen.\ Rel.\ Grav.\  {\bf 38} (2006) 1797
 \href{http://arxiv.org/abs/gr-qc/0605049}{ [arXiv:gr-qc/0605049]}.

\bibitem{VisserBook}
  M.~Visser,
  \emph{Lorentzian wormholes: From Einstein to Hawking},
  Woodbury, USA: AIP (1995). 

\bibitem{KRW}
  B.~S.~Kay, M.~J.~Radzikowski and R.~M.~Wald,
  ``Quantum field theory on spacetimes with a compactly generated Cauchy
  horizon,''
  Commun.\ Math.\ Phys.\  {\bf 183}, 533 (1997)
  \href{http://arxiv.org/abs/gr-qc/9603012}{[arXiv:gr-qc/9603012]} .


\bibitem{Nordstrom} G.~Nordstr\"om,
\emph{Zur Theorie der Gravitation vom Standpunkt des Relativit�sprinzip},
Annalen der Physik,
42: 53354. 
 \\
  J. Norton {\it Einstein, Nordstr\"om and the early Demise of Lorentz-covariant, Scalar Theories of Gravitation}, Archive for History of Exact Sciences, 45 (1992),  \href{http://www.pitt.edu/~jdnorton/papers/Nordstroem.pdf}{http://www.pitt.edu/~jdnorton/papers/Nordstroem.pdf}.
\\
  C.~W.~Misner, K.~S.~Thorne and J.~A.~Wheeler,
\emph{ Gravitation},
 San Francisco (1973)



\bibitem{Will}
  C.~M.~Will,
  \emph{The confrontation between general relativity and experiment},
  Living Rev.\ Rel.\  {\bf 9}, 3 (2005)
  \href{http://www.livingreviews.org/lrr-2006-3}{http://www.livingreviews.org/lrr-2006-3};
  \href{http://arxiv.org/abs/gr-qc/0510072}{[arXiv:gr-qc/0510072]}.

\bibitem{smolinemqm}
  F.~Markopoulou and L.~Smolin,
  \emph{Quantum theory from quantum gravity,}
  Phys.\ Rev.\  D {\bf 70}, 124029 (2004)
  \href{http://arxiv.org/abs/gr-qc/0311059}{[arXiv:gr-qc/0311059]} .

\bibitem{stefano}   C.~Barcelo, S.~Liberati and M.~Visser,
  \emph{Analogue gravity},
  Living Rev.\ Rel.\  {\bf 8} (2005) 12
  \href{http://arxiv.org/abs/gr-qc/0505065}{[arXiv:gr-qc/0505065]}.

\bibitem{lorenzo}
 F.~Girelli, S.~Liberati and L.~Sindoni,
  {\it Gravitational dynamics in Bose Einstein condensates},
  Phys.\ Rev.\  D {\bf 78} (2008) 084013 \href{http://xxx.lanl.gov/abs/0807.4910}{[gr-qc/0807.4910]}.

\bibitem{jacobson}
  T.~Jacobson,
  \emph{Thermodynamics of space-time: The Einstein equation of state},
  Phys.\ Rev.\ Lett.\  {\bf 75}, 1260 (1995)
  \href{http://arxiv.org/abs/gr-qc/9504004}{[arXiv:gr-qc/9504004]};\\
  C.~Eling, R.~Guedens and T.~Jacobson,
  \emph{Non-equilibrium Thermodynamics of Spacetime},
  Phys.\ Rev.\ Lett.\  {\bf 96}, 121301 (2006)
  \href{http://arxiv.org/abs/gr-qc/0602001}{[arXiv:gr-qc/0602001]}.


\bibitem{conrady}
  F.~Conrady and L.~Freidel,
  ``On the semiclassical limit of 4d spin foam models,''
  Phys.\ Rev.\  D {\bf 78}, 104023 (2008)
  \href{http://arxiv.org/abs/0809.2280}{[arXiv:0809.2280]}.
  
\bibitem{oriti}
  D.~Oriti,
  \emph{Group field theory as the microscopic description of the quantum spacetime
  fluid: a new perspective on the continuum in quantum gravity,}
  \href{http://arxiv.org/abs/gr-qc/0710.3276}{[arXiv:0710.3276]}.

\bibitem{lorenzo1}
 F.~Girelli, S.~Liberati and L.~Sindoni,
  {\it Emergence of Lorentzian signature and scalar gravity},
  Phys.\ Rev.\ D {\bf 79}
    \href{http://xxx.lanl.gov/abs/0806.4239}{[gr-qc/0806.4239]}.

\bibitem{ignatowsky}   S.~Liberati, S.~Sonego and M.~Visser,
\emph{Faster-than-c signals, special relativity, and causality},
  Annals Phys.\  {\bf 298} (2002) 167
   \href{http://arxiv.org/abs/gr-qc/0107091}{[arXiv:gr-qc/0107091]}.
   W.A.~von Ignatowsky,
\emph{Einige allgemeine Bemerkungen zum Relativit\"atsprinzip},
Verh. Deutsch. Phys. Ges. 12, 788 (1910);
\emph{Einige allgemeine Bemerkungen zum Relativit\"atsprinzip},
Phys. Zeitsch. 11, 972
(1910);
\emph{Das Relativit\"atsprinzip},
Arch. Math. Phys. 3 (17), 1; (18), 17 (1911);
\emph{Eine Bemerkung zu meiner Arbeit Einige allgemeine Bemerkungen zum Relativitatsprinzip},
Phys. Zeitsch. 12, 779 (1911).
\bibitem{brisbane} Pictures taken from \href{http://www.bris.ac.uk/Depts/Chemistry/MOTM/diamond/diamond.htm}{http://www.bris.ac.uk/Depts/Chemistry/MOTM/diamond/diamond.htm}.


  \bibitem{fokker}
  A.~Einstein and A.~D.~Fokker,
  Nordstr\"om's Theory of Gravitation from the Point of View of the Absolute Differential Calculus,
  Annalen Phys.\  {\bf 44}, 321 (1914)
  [Annalen Phys.\  {\bf 14}, 500 (2005)].

\bibitem{sonego}    H.~Westman and S.~Sonego,
  \emph{Coordinates, observables and symmetry in relativity},
  \href{http://arxiv.org/abs/gr-qc/0711.2651}{[arXiv:0711.2651]}.

\bibitem{WW}
  S.~Weinberg and E.~Witten,
  \emph{Limits On Massless Particles},
  Phys.\ Lett.\  B {\bf 96}, 59 (1980).

\bibitem{Jenkins}
  A.~Jenkins,
  \emph{Topics in particle physics and cosmology beyond the standard model},
  \href{http://arxiv.org/abs/hep-th/0607239}{[arXiv:hep-th/0607239]}.



\end{thebibliography}
\end{document}